# Spline-based Interface Modeling and Optimization (SIMO) for Surface Tension and Contact Angle Measurements

Karan Jakhar[a,b], Ashesh Chattopadhyay[a,c], Atul Thakur[a], and Rishi Raj[a, 1]

[a]Department of Mechanical Engineering, Indian Institute of Technology Patna, Patna, Bihar 801103, India

[b]Currently with: J. Mike Walker '66 Department of Mechanical Engineering, Texas A&M University, College Station, TX 77843, USA

[c]Currently with: Department of Mechanical Engineering, Rice University, Houston, TX 77005,, USA

## ABSTRACT

Surface tension and contact angle measurements are fundamental characterization techniques relevant to thermal and fluidic applications. Drop shape analysis techniques for the measurement of interfacial tension are powerful, versatile and flexible. Here we develop a Spline-based Interface Modeling and Optimization (SIMO) tool for estimating the surface tension and the contact angle from the profiles of sessile and pendant drops of various sizes. The employed strategy models the profile using a vector parametrized cubic spline which is then evolved to the eventual equilibrium shape using a novel thermodynamic free-energy minimization-based iterative algorithm. We perform experiments to show that in comparison, the typical fitting-based techniques are very sensitive to errors due to image acquisition,

[1] Corresponding Author. Telephone: +91 612 302 8166. E-mail address: rraj@iitp.ac.in {Rishi Raj}
Address: Room 113, Block III, Department of Mechanical Engineering, Indian Institute of Technology Patna, Bihta, Bihar 801103, India

digitization and edge detection, and do not predict the correct surface tension and the contact angle values. We mimic these errors in theoretical drop profiles by applying the Gaussian noise and the smoothing filters. We then demonstrate that our optimization algorithm can even drive such inaccurate digitized profiles to the minimum energy equilibrium shape for the precise estimation of the surface tension and the contact angle values. We compare our scheme with software tools available in public domain to characterize the accuracy and the precision of SIMO.

***Keywords****: Surface Tension; Contact Angle; Drop Shape Analysis; Cubic Spline; Spline-based Interface Modeling (SIM), Spline-based Interface Modeling and Optimization (SIMO)*

# 1. Introduction

Interfacial tension is a concept of fundamental importance in surface science, describing the phenomenon as diverse as formation, shape, and stability of liquid drops [1,2]. Moreover, contact angle provides information about the chemical composition, roughness and heterogeneity of the solid surface [3–8]. This has enabled a constant evolution of liquid-vapor interfacial tension and contact angle measurement techniques [2,9].

Contact angles may be directly determined using a goniometer by simply aligning the tangent to the side profile of drop at the three-phase contact point. The major limitation of this method is the constraint which requires the user to dispense ultra-small droplets such that the effect of gravity can be neglected and easy to implement circular/spherical cap approximation for droplet profile may be applied [10]. Conversely, the polynomial fitting approach often used to determine contact angles of large drops (where the effect of gravity on droplet shape deviation from circular/spherical cap cannot be neglected) are highly sensitive to the degree of the polynomial and the number of coordinate points [11,12]. 'DropSnake', is a spline-based approach to estimate the contact angles. The elastic property of spline (snake) links the local nature of contact angle to the global contour of the drop to reveal the contact angle. The global model allows for finding the symmetry in the image utilizing its reflection, which can enhance the detection of drop's baseline and tilt angle. DropSnake has proven to be robust and is openly available for use [13].

Tensiometry, the measurement of the interfacial tension between fluid phases, directly probes the competition between intermolecular forces that give rise to interfacial tension, and long-range gravitational or externally applied forces that deform the liquid interface. Out of the numerous tensiometry methods devised [2], drop shape analysis based sessile and pendant tensiometry methods have been popular to determine fluid-fluid interfacial tension from the shapes of drops or bubbles [14–22]. Interfacial tension is usually back-calculated by estimating and matching the gravitational deformation of a drop or a bubble to the solution of the Young-Laplace equation [3,23]. For example, the ADSA (Axisymmetric Drop Shape Analysis) method requires solving the Young-Laplace equation by numerical integration. After the discretization of the contour of a drop on an image, it searches for the best Laplace profile (theoretical curve obtained by solving the Young-Laplace equation) that corresponds to this contour. One may then obtain an accurate contact angle as well as surface tension. The accuracy of ADSA crucially depends on the quality of captured profile and edge detection [24].

The accuracy of surface tension estimation is reduced at low Bond numbers ($Bo = \rho g L^2/\gamma$) due to fundamental physical limitations [24]. Here $\rho$ and $\gamma$ are the density and surface tension of fluid, $g$ is acceleration due to gravity, and $L$ is the characteristic dimension of the drop defined as $L = (3V/4\pi)^{1/3}$ where $V$ is the volume of the drop. This limitation is often overcome using modifications in tensiometry techniques where an extra interface is included in the form of a solid surface or particle [19,25–27]. Berry $et.$ $al.$ [28] found drop volume as an overriding criterion in determining measurement precision contrary to the literature reports [24,29]. Berry $et$ $al.$ [28] introduced a new non-dimensional quantity, the Worthington number ($Wo$), to account for this volume effect, where large value of $Wo$ indicates greatest precision ($Wo$ values scale from 0 to 1). In addition, Berry $et.$ $al.$ [28] developed an open-source python program 'OpenDrop' to facilitate accurate estimation of surface and interfacial tension of pendant drops.

Our scheme to determine fluid-fluid interfacial tension and contact angle from the shape of axisymmetric liquid-vapor interface; i.e., from sessile as well as pendant drops is developed on a drop shape prediction and analysis scheme from our previous work [30]. In our previous work, we developed a vector-valued parametrized cubic spline-based representation for modeling liquid-vapor interface. It was equipped with a thermodynamic free energy minimization-based heuristic to provide a geometric solution to the Young-Laplace equation.

We predicted the shapes of drops under the action of gravitational (sessile and pendant drops) and centrifugal forces. We extended our Spline-based Interface Modeling (SIM) approach to perform inverse analysis, i.e., analyze a given drop profile to predict the interfacial properties. In this work, we use experimental images to show that this fitting-based drop shape analysis technique (SIM) is very sensitive to minor deviations (due to image acquisition, edge detection, experimental setup, etc.) from the theoretical drop profile corresponding to the solution of Young-Laplace equation and does not predict correct surface tension and contact angle values. We develop an optimization algorithm for SIM (SIM-O: Spline-based Interface Modeling and Optimization) to address these limitations and drive the inaccurate digitized profiles to the minimum energy equilibrium shape for the precise estimation of surface tension and contact angle values. We believe that SIMO can serve as a powerful drop shape analysis tool to supplement surface tension and contact angle measurement research.

## 2. Modeling

SIMO is equipped with thermodynamic free energy minimization algorithm from our previous work [30]. It models equilibrium, i.e., no liquid flow/circulation inside the drop. For example, when no external force is acting on the drop ($Bo = 0$), it is well known to adopt a spherical cap shape (constant curvature $\chi$) such that the internal gauge pressure ($-2\gamma\chi_{mean}$) is a constant. Here $\chi_{mean} = \frac{\chi_1 + \chi_2}{2}$ is the mean of principal curvatures $\chi_1$ and $\chi_2$. $\chi_1$ and $\chi_2$ are defined below in Eqs (1) and (2). The case of a large drop under the action of gravity ($Bo \neq 0$) implies that the internal pressure follows hydrostatics, i.e., the equivalent mean curvature ($\chi_{eq} = \chi_{mean} - \rho g h / 2\gamma$) is reduced to a constant.

### 2.1 Spline-based representation

We represent the liquid-vapor interface of drop as a vector parametrized cubic spline $S = [x(u) \quad y(u)]$, where $u \in [0, 1]$ represents the normalized spline parameter. The spline-based parametrized representation makes it convenient to compute the curvature, surface area, volume and contact angle of an axisymmetric drop profile. A commercial software package MATLAB® [31] is used to perform all computations. Inverse tangent of $dy(u)/dx(u)|_{u=0,1}$ is evaluated to calculate left and right contact angles, where start and end points (contact points) of parametrized spline are represented by $u$ of 0 and 1 respectively. The principal curvatures across the spline for all points are evaluated as follows:

$$\chi_1 = \frac{1}{r_1} = \frac{x'y'' - y'x''}{(x'^2 + y'^2)^{3/2}} \qquad (1)$$

and,

$$\chi_2 = \frac{1}{r_2} = \frac{y'}{x(x'^2 + y'^2)^{1/2}} \qquad (2)$$

where prime ($'$) represents the derivative *w.r.t.* the parameter $u$.

It can be seen from Eqs. (1) and (2) that $\chi_{mean}$ is defined –ve for concave outward and +ve for concave inward interfaces. The perimeter defined as $P_{lv} = \int_{u=0}^{u=1} dS$ is evaluated and Pappus's centroid theorem [32] facilitates the calculation of surface area and volume of the drop. The details of implementation of a cubic spline for modeling drops can be found in our previous work [30].

## 2.2 SIM (Spline-based Interface Modeling)

We use spline-based representation and our thermodynamic based free energy minimization approach to perform an inverse analysis on drop shapes, i.e., back-calculate $Bo$ from the profile of fluid-fluid interface of an axisymmetric drop. Inverse analysis starts with digitizing the given drop profile to obtain coordinates of all points along the liquid-vapor interface. After representing the acquired drop profile coordinates using spline, the mean curvature ($\chi_{mean}$) is evaluated for all points. We then assume a value of $Bo$ to estimate the equivalent curvature ($\chi_{eq}$) at all points, which is further used to estimate the standard deviation in the equivalent curvature ($\sigma_{\chi_{eq}}$) for this assumed value of $Bo$. The above-mentioned procedure is repeated refining the guessed value of $Bo$, which is allowed to span the large parametric space of the drop surface. $\sigma_{\chi_{eq}}$ and $Bo$ are plotted against each other and the $Bo$ corresponding to the minimum value of $\sigma_{\chi_{eq}}$ is identified, i.e., a fairly constant $\chi_{eq}$ implying the minimum energy equilibrium state [30].

An example calculation for the normalized drop shapes obtained by our drop shape prediction algorithm [30] with unit volume and corresponding to $Bo = -0.24, 0, 0.5$ and contact angle, $\theta = 133°$ is shown in Fig. 1. It is evident that the standard deviation of equivalent curvature ($\sigma_{\chi_{eq}}$) is minimized at $Bo = -0.24, 0, 0.5$ respectively, identical to the actual $Bo$ value of drop shapes. The corresponding contact angle values ($\theta$) also match the actual contact angle values.

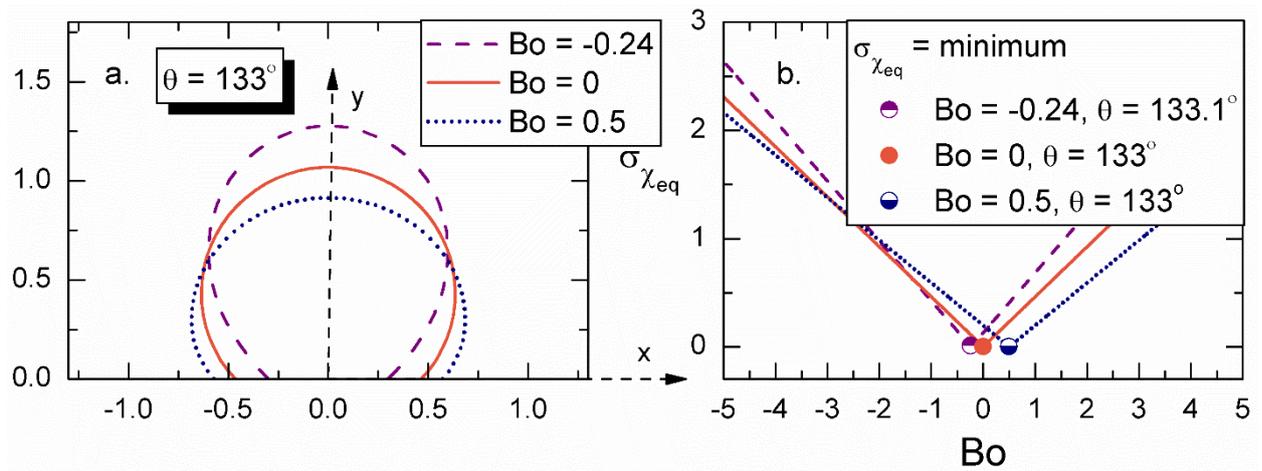

**Fig. 1.** (a) Axisymmetric drop shapes for $Bo = -0.24, 0, 0.5$ with $\theta = 133°$. (b) $\sigma_{\chi_{eq}}$ approaches the minimum value ($\approx 0$) at $Bo = -0.24, 0, 0.5$ for $\theta \approx 133°$.

**2.3 Experiment**

SIM [30] works well for theoretical drop profiles, however, when it comes to experimental shapes, the digitized experimental drop profiles are equipped with numerous errors including but not limited to errors due to image acquisition, edge detection, and pixelization, among others. Hence, we analyse digitized experimental drop profile in this section to assess the accuracy of SIM with experimental images.

A typical pendant drop tensiometry experiment was set-up to capture pendant drop profiles. The experimental apparatus includes a needle, syringe pump (Cole-Palmer, WW-74905-04), a high-speed camera (Phantom v7.3), and a light source, similar to those used in the literature [28]. Millipore water was dispensed with constant volume addition rate and images were captured at uniform time intervals. Volume addition rate was kept low enough such that the process can be assumed to be quasi-steady/equilibrium. Images were captured using the high-speed camera with a resolution of $800 \times 600$ pixels. Please note that the experimentally captured images and a dataset of drops images created by employing droplet shape prediction algorithm [30] for all analyses in this manuscript are digitized using canny edge detector [33] of Image Processing Toolbox of MATLAB [31]. Values of the threshold and the standard deviation of the Gaussian filter were appropriately decided. Contact points were manually identified.

A sample image of a water drop from our experiment is shown in Fig. 2a (the results of analysis for full cycle until detachment are discussed later in Fig. 8). A digitized profile of the drop image is obtained and made axisymmetric. For a digitized profile with $N$ ($N =$ even) discretized coordinates, the $y$ coordinates of the symmetric points ($[1, N], [2, N-2], [3, N-3]$, ...) are assigned a value equal to their mean ($y_{i,N} = (y_i + y_N)/2$), while the $x$ coordinates of the symmetric points are shifted along $x$ axis such that their mean is equal to 0 ($x_{i,N} = x_{i,N} \pm (x_i + x_N)/2$) making the drop profile axisymmetric. If $N$ is odd, the $y$ coordinate of apex point or $(N + 1)/2^{\text{th}}$ coordinate is kept unchanged and the value of the $x$ coordinate is made 0. Coordinates other than apex coordinate follows process similar to that of even number of coordinates ($N = $ even).

We then apply SIM on the obtained axisymmetric shape to estimate the $Bo$ corresponding to the minimum value of $\sigma_{\chi_{eq}}$ (Fig. 2b). Please note that the drop shape is shown to the scale while other parameters including $\chi_{eq}$, $\sigma_{\chi_{eq}}$, among others are represented for normalized drop shape (normalized to unit volume) for all analyses in this manuscript. Contrary to the theoretical shape in Fig. 1b, the minimum value of $\sigma_{\chi_{eq}} \neq 0$. Moreover, the estimated $Bo = -0.251$ and $\theta = 128.1°$ corresponding to the minimum $\sigma_{\chi_{eq}}$ for the digitized drop shape. Please note that $\theta$ for pendant drop with needles is the angle between the interface and an assumed flat surface as shown in Fig. 2a. This definition of $\theta$ will be applied in the manuscript for all pendant drops suspended from a needle. The usual definition of contact angle $\theta$ will apply for all other drops in further discussions in this manuscript. Surface tension ($\gamma = 68.67$ mN/m for digitized drop image) is easily calculated using Eqs. (1) and (2) since $Bo$ and other drop parameters are known ($\gamma = \rho g L^2 / Bo$). The calculated surface tension value is significantly off from the theoretical value of surface tension for water ($\gamma = 72.75$ mN/m [34]) at 20 °C.

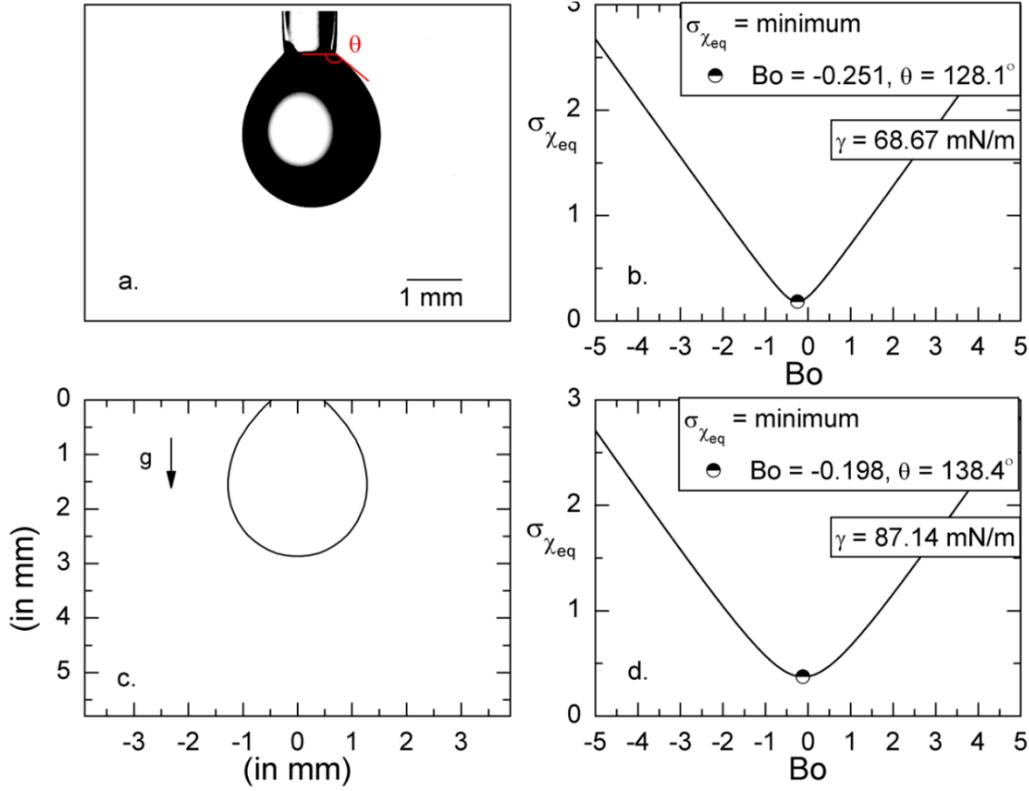

**Fig. 2.** (a) Experimental image of a pendant drop. (b) $\sigma_{\chi_{eq}}$ approached minimum value at $Bo = -0.251, \theta = 128.1°$ for the digitized experimental drop shape and the corresponding $\gamma = 68.67$ mN/m. (c) Pixelized shape of drop mimicking the errors in image acquisition due to discretization. (d) $\sigma_{\chi_{eq}}$ approaches minimum value at $Bo = -0.198, \theta = 138.4°$ for the pixelized drop shape and the corresponding $\gamma = 87.14$ mN/m.

We next use our droplet shape prediction algorithm [30] to obtained a drop profile with volume and the base diameter (needle diameter) same as that of the experimental drop shape in Fig. 2a. The values of density and surface tension are assumed to be equal to the literature values of water at 20 °C, i.e. 998 kg/m³ [35] and 72.75 mN/m [34] respectively; $\theta = 133°$ for the profile. Since this is a perfect theoretical drop shape (similar to those in Fig. 1a) without any error, SIM accurately predicts the correct value of $Bo = -0.237$ and $\theta = 133°$. And as was shown in Fig. 1b, the minimum value $\sigma_{\chi_{eq}}$ also approaches 0 at this value of $Bo$.

We next partition the window of known length (length identical to the widow length of captured image in Fig. 2a) containing the drop shape into $800 \times 600$ pixels. This mimics the grid of pixels in an image wherein every small square formed in this grid is analogous to a pixel of the experimentally captured image of the same resolution. The coordinates of the theoretical (exact) drop profile lying in any of those small squares are rounded off to the centre coordinate of the individual small squares. This process pixelized the drop shape and is shown in Fig. 2c. The minor changes in the coordinates of profile before (theoretical) and after pixelization (mimicking experiments) can be found in supporting section S1. Similar to Fig. 2b, the minimum value of $\sigma_{\chi_{eq}}$ is neither 0, nor the estimated values of Bond number, contact angle, and hence the surface tension are accurate (SIM computes $Bo = -0.198$, $\theta = 138.4°$, and $\gamma = 87.14$ mN/m for digitized drop shape). This suggests that the fitting-based drop shape analysis technique (SIM) is very sensitive to such practical limitations and should be supplemented with some post-processing for the accurate estimation of the contact angle and surface tension values.

## 2.4 Spline-based Interface Modeling and Optimization (SIMO)

We now tackle the inaccurate drop profiles (black solid profile in Fig. 2c/Fig. 3a) by introducing an optimization routine to SIM and abbreviate it as SIMO/SIM-O (Spline-based Interface Modeling and Optimization). We utilize the Pinning (PN) perturbation operator introduced in our previous work [30], which when used iteratively (alternate +PN and –PN) drives an initialized/inaccurate drop shape (the black profile in supporting Video S1a and S2a) with non-uniform pressure distribution across liquid-vapor interface ($\chi_{eq} \neq$ constant; black curve in supporting Video S1b and S2b) to the final equilibrium shape corresponding the assumed value of $Bo$ (magenta profile in supporting Video S1a for $Bo = -0.15$ and Video S2a for $Bo = -0.237$) without moving the contact line. In +PN, a small volume $\Delta V$ of liquid is added to the drop profile at the location of the highest instantaneous equivalent curvature ($\chi_{eq}$; filled blue square in supporting Video S1 and S2). It is then followed by –PN, i.e., a PN operator where same amount of liquid is removed ($-\Delta V$) from the location of the lowest instantaneous equivalent curvature (filled red circle in supporting Video S1 and S2) such that the overall volume is maintained constant. Above mentioned two operators +PN and –PN in turn imitate the natural tendency of the fluid to flow from the location of higher internal pressure (filled red circle in supporting Video S1 and S2) to lower pressure (filled blue square in supporting Video S1 and S2). Iterative use of these two operators eventually reduces the $\chi_{eq}$ to a constant (magenta curve in supporting Video S1b and S2b) driving the erroneous experimental shapes to the perfect theoretical drop shape for the specified volume and the assumed value of Bond number, i.e., also a geometrical solution of Young-Laplace equation. Please refer to our previous work for further details on the implementation of these operators [30].

$Bo$ corresponding to the minimum value of $\sigma_{\chi_{eq}}$ is identified by evaluating a drop profile in SIM and an interval encapsulating this $Bo$ value is taken as the range of guess values of $Bo$ for further analysis in SIMO. For example, we operate SIMO on the experimental drop shape from Fig. 2a with a guess value of $Bo = -0.15$. Error introduced drop shape (solid black curve in Fig. 3a and black profile in supporting Video S1a) is iteratively evolved using +PN and –PN operators (blue profile in supporting Video S1a) until the equilibrium position ($\sigma_{\chi_{eq}} \approx 0$, dashed green curve in Fig. 3a and magenta profile in supporting Video S1a) is attained. The equilibrium shape is also the theoretical shape for the guess value of $Bo$. Solid black curve in Fig. 3b is a subset of Fig. 2b showing the values of $\sigma_{\chi_{eq}}$ for the profile in Fig. 2a. Please note that unlike SIM, $\sigma_{\chi_{eq}}$ eventually reduces to 0 (dashed curve in Fig. 3b) upon operating SIMO. The upward facing filled blue and red triangle in Fig. 3b and supporting Video S1c represents value of $\sigma_{\chi_{eq}}$ for initial and final shapes respectively for guess value of $Bo = -0.15$.

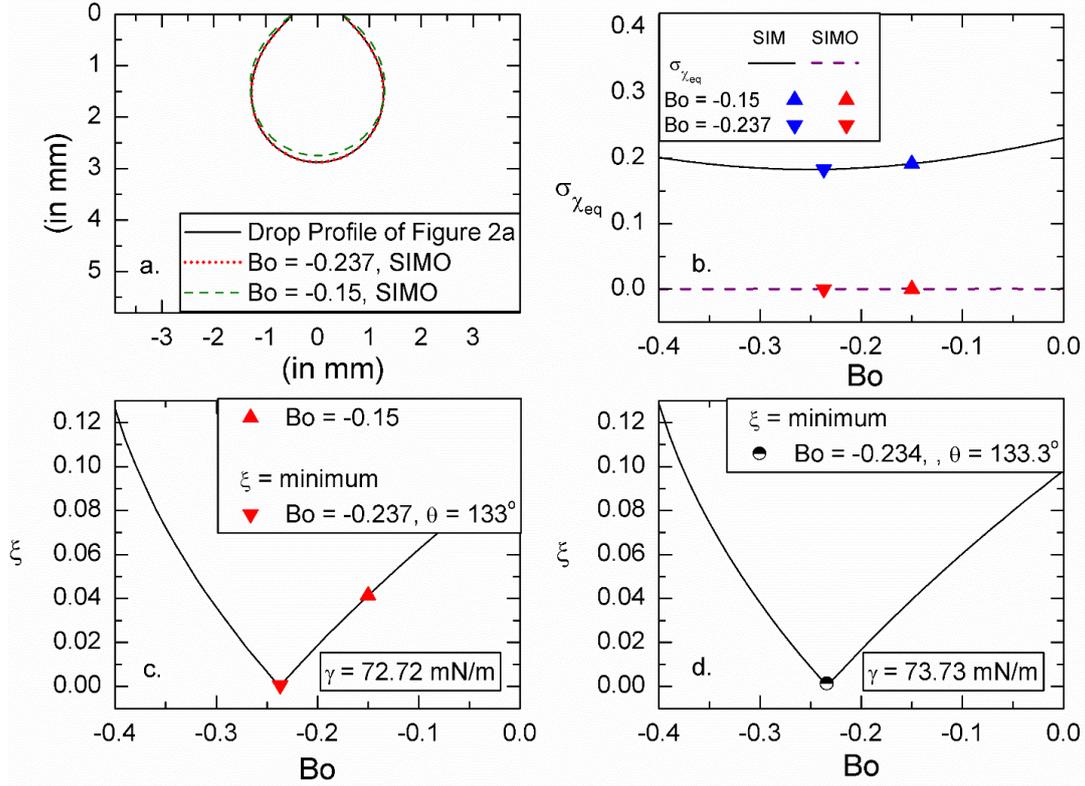

**Fig. 3.** (a) Evolution of digitized drop profile from Fig. 2a (solid black line) to final equilibrium position for $Bo = -0.15$ (dashed green line) and $Bo = -0.237$ (dotted red line) by SIMO. (b) $\sigma_{\chi_{eq}}$ for digitized drop profile in Fig. 2a (solid black line; subset of Fig. 2b) after evolution to final equilibrium position (dashed line) by SIMO. (c) $\xi$ approaches minimum value ($\xi \approx 0$) at $Bo = -0.237, \theta = 133°$ (downward facing filled red triangle) corresponding to $\gamma = 72.72$ mN/m for digitized experimental drop shape, and, (d) $Bo = -0.234, \theta = 133.3°$ (half-filled black circle) corresponding to $\gamma = 73.73$ mN/m for pixelized drop shape after running SIMO on digitized drop profile in Fig. 2a and pixelized drop profile in Fig. 2c respectively.

Since $\sigma_{\chi_{eq}} \approx 0$ for the final equilibrium shapes for all guess values of $Bo$ (dashed curve in Fig. 3b), a new criterion, Root Mean Square Deviation ($\xi$) is introduced to identify the correct $Bo$ for the drop shape.

$$\xi = \sqrt{\sum_{i=1}^{N} \frac{(y_i - y_i^{eq})^2 + (x_i - x_i^{eq})^2}{N}} \qquad (3)$$

Here $x_i^{eq}$ and $y_i^{eq}$ are coordinates of drop shape at final equilibrium shape (magenta profile in supporting Video S1a and S2a). For example, the upward facing filled red triangle shows $\xi$ for the guess value of $Bo = -0.15$ in Fig. 3c. $\xi$ between the initial shape and the equilibrated shape is obtained and the $Bo$ value corresponding to the minimum value of $\xi$ is identified as the $Bo$ value of the drop. The equilibrium shape (dotted red curve in Fig. 3a) corresponding to the minimum value of $\xi$ (downward facing filled red triangle in Fig. 3c) for the guess value of $Bo = -0.237$ has the minimum deviation in shape with respect to the original profile. Supporting Video S2 presents the evolution of drop profile from Fig. 2a (black profile) to final equilibrium shape (magenta profile) for this correct guess value of $Bo = -0.237$.

Having estimated the correct value of $Bo$, the surface tension is determined using the values of the acceleration due to gravity, density of the fluid (or difference in density of two fluids for fluid-fluid interface) and the volume ($\gamma = \rho g L^2 / Bo$). Contact angle can then be easily obtained from the drop shape at equilibrium as explained in the aforementioned Section 2.1 of spline-based representation. Similar procedure is repeated for the pixelized drop profile in Fig. 2c and the subsequent $\xi$ is shown in Fig. 3d. The obtained value of $Bo = -0.234$ and the resulting value of $\gamma = 73.73$ mN/m for the pixelized drop shape is once again close to the reported surface tension of water ($\gamma = 72.75$ mN/m at 20 °C). Additionally, $\theta = 133.3°$ for pixelized drop profile is also close to the actual $\theta = 133°$.

## 3. Results and Discussion

We now use the Spline-based Interface Modeling and Optimization (SIMO) scheme to analyze various sessile ($Bo > 0$), pendant ($Bo < 0$), and spherical ($Bo = 0$) drop shapes to gauge its limitations. We analyze a few digitized drop images and compare the accuracy of predictions from SIMO with the tools available in the public domain.

### 3.1 Digitized Drop Images

We created a dataset of drops images by employing droplet shape prediction algorithm from our previous work [30]. The dataset includes a range of sessile, pendant and spherical drops with various contact angle ($\theta$) values. In order to mimic the errors in digitized experimental drops, the drop shapes were subsequently altered by means of a Gaussian smoothing filter and Gaussian noise filter. Smoothing was linearly incremented in five steps up to the smoothing standard deviation of 5. Similarly, the noise was incremented in five steps up to noise relative variance of 0.25. Fig. 4 shows a sample of altered drop images.

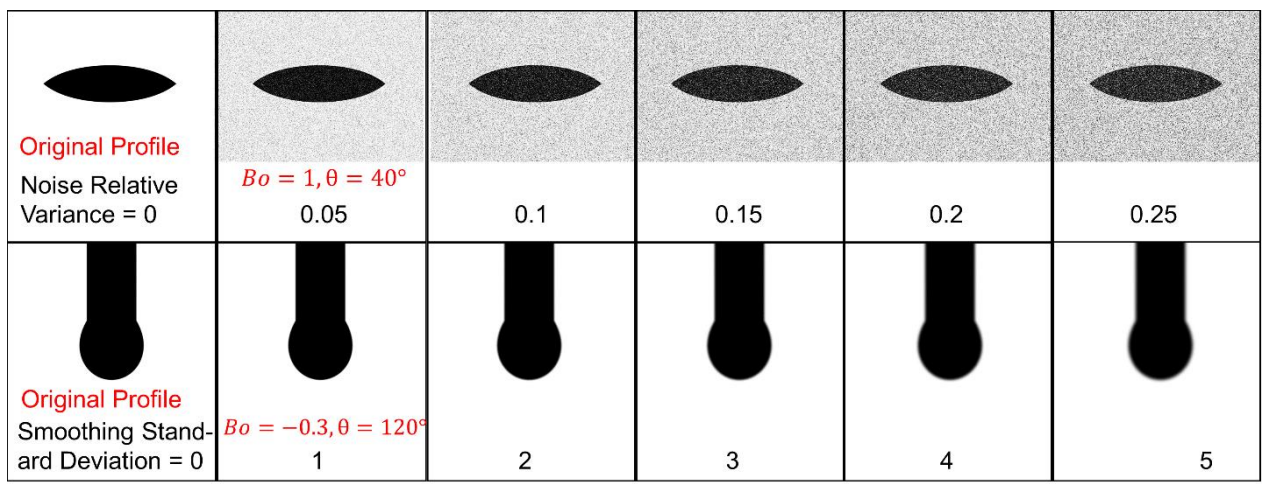

**Fig. 4.** Samples of drop profile before and after application of Gaussian noise and smoothing filters.

We now report the results of analysis on these altered digitized pendant drop images with actual $Bo$ and $\theta$ values mentioned in Table 1 (Test #1 to #6). Resulting computed values of $Bo$ by SIM and SIMO are shown in Fig. 5. Solid black line represents good agreement between the estimated and actual values of $Bo$. Span of results along the $y$ axis and their distance from the black line represents the inaccuracy in prediction. For example, SIM predicts $Bo = 2.84$ while SIMO predicts $Bo = 1.05$ for the theoretical drop image with $Bo = 1$. Similar analysis to predict $\theta$ for theoretical pendant drop images is shown in Fig. 6 for actual $Bo$ and $\theta$ values mentioned in Table 1 (Test #7 to #12). Solid black line represents good agreement between the estimated and actual values of $\theta$. The trend in Fig. 6 is similar to that in Fig. 5 with precise and accurate results from SIMO while grossly deviating results for SIM. For example, SIM predicts $\theta = 108°$ and SIMO predicts $\theta = 119°$ for theoretical drop image with actual $\theta = 120°$ and noise relative variance = 0.05. These results signify the robustness of SIMO for measuring $Bo$ (and subsequently surface tension) and contact angles. Additionally, we can also acknowledge its ability to address the inaccuracies with blurry and noisy images.

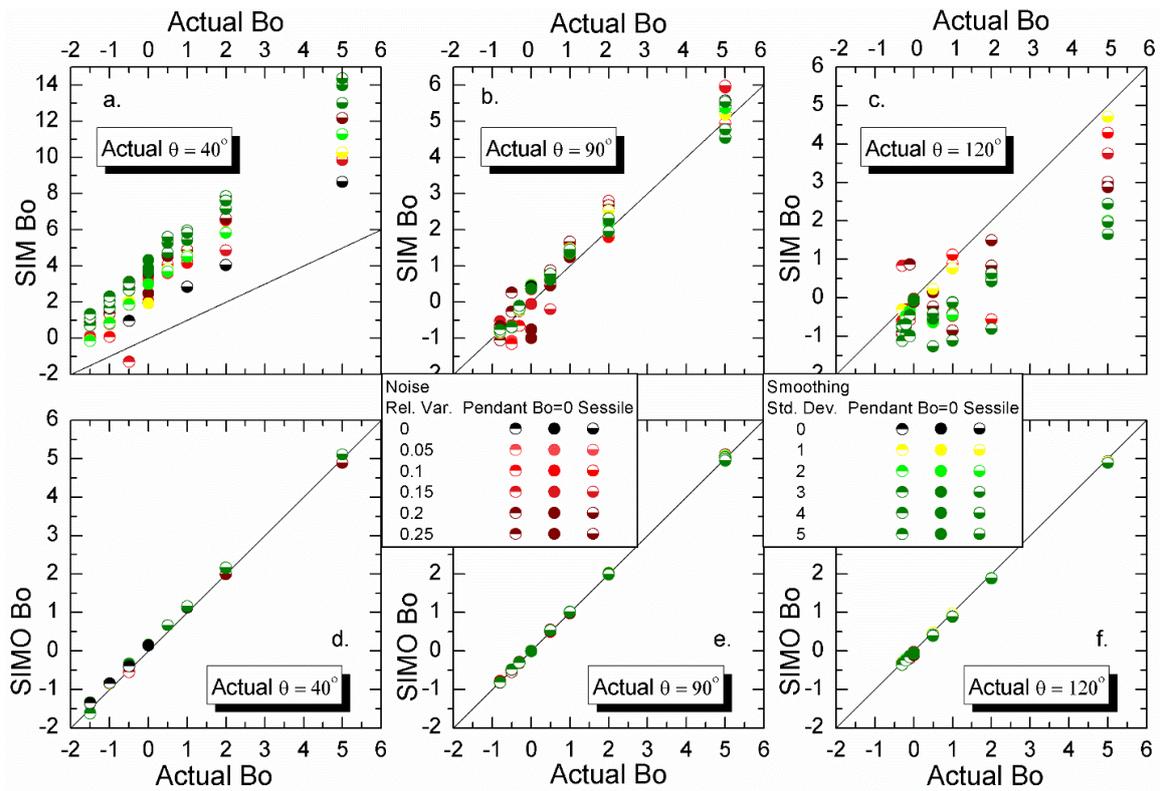

**Fig. 5.** Predicted $Bo$ for drop profiles with various $Bo$ values before and after application of Gaussian noise and smoothing filter for (a) Test #1, (b) Test # 2, (c) Test # 3, (d) Test # 4, (e) Test # 5, and, (f) Test # 6.

**Table 1.** Test matrix for the analyses performed in Figs. 5-8.

| Test # | Actual $Bo$ | Actual $\theta$ | Scheme | Result |
|---|---|---|---|---|
| 1 | $-1.5, -1, -0.5, 0, 0.5, 1, 2, 5$ | 40° | SIM | Fig. 5a |
| 2 | $-0.8, -0.5, -0.3, 0, 0.5, 1, 2, 5$ | 90° | SIM | Fig. 5b |
| 3 | $-0.3, -0.2, -0.1, 0, 0.5, 1, 2, 5$ | 120° | SIM | Fig. 5c |
| 4 | $-1.5, -1, -0.5, 0, 0.5, 1, 2, 5$ | 40° | SIMO | Fig. 5d |
| 5 | $-0.8, -0.5, -0.3, 0, 0.5, 1, 2, 5$ | 90° | SIMO | Fig. 5e |
| 6 | $-0.3, -0.2, -0.1, 0, 0.5, 1, 2, 5$ | 120° | SIMO | Fig. 5f |
| 7 | $-0.2$ | | SIM | Fig. 6a |
| 8 | $0$ | | SIM | Fig. 6b |
| 9 | $2$ | | SIM | Fig. 6c |
| 10 | $-0.2$ | 40°, 90°, 120° | SIMO | Fig. 6d |
| 11 | $0$ | | SIMO | Fig. 6e |
| 12 | $2$ | | SIMO | Fig. 6f |
| 13 | $-0.6$ to $0$ in an interval of 0.05 | 100° | SIMO and OpenDrop [28] | Fig. 7a |
| 14 | $-0.4$ to $0$ in an interval of 0.05 | 110° | | |
| 15 | $-0.35$ to $0$ in an interval of 0.05 | 120° | | |
| 16 | $-0.2$ to $0$ in an interval of 0.05 | 130° | | |
| 17 | $-0.15$ to $0$ in an interval of 0.05 | 140° | | |
| 18 | -0.5, 0, 1 | 10° to 90° in an interval of 5° | SIMO and DropSnake [13] | Fig. 7b |
| 19 | -0.1 | 95° to 145° in an interval of 5° | | |
| 20 | -0.05 | 150°, 155° | | |
| 21 | 0, 1 | 95° to 170° in an interval of 5° | | |

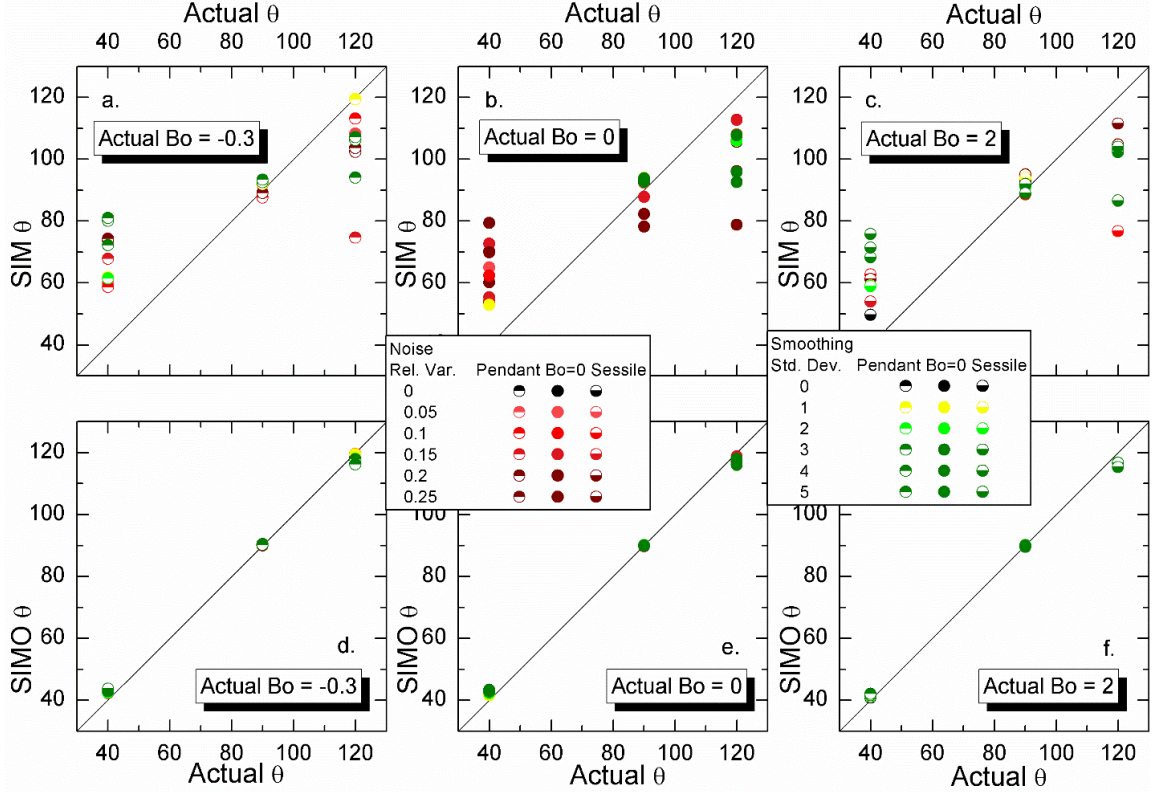

**Fig. 6.** Predicted $\theta$ for drop profiles with various $\theta$ values before and after application of Gaussian noise and smoothing filter for (a) Test # 7, (b) Test # 8, (c) Test # 9, (d) Test # 10, (e) Test # 11, and, (f) Test # 12.

## 3.2 Comparison with other Algorithms

We next present the results of analysis on a few more unaltered digitized drop images and compare the results of SIMO with the openly available tools, 'OpenDrop'[28], a surface tension measurement algorithm for pendant drops, and, 'DropSnake' [13], a contact angle measurement algorithm.

A set of unaltered digitized drop images were selected for $Bo$ measurement as mentioned in Table 1 (Test #13 to #17). The drop shapes used for this analysis are similar to the original profile of pendant drop with needle in Fig. 4. The computed values of $Bo$ by SIMO and OpenDrop are presented in red and blue, respectively (Fig. 7a). Solid black line suggests a good agreement between the predicted and actual values of $Bo$. $Bo$ values predicted by OpenDrop and SIMO demonstrate comparable accuracy over a wide range of Bo and $\theta$ values.

Similarly, a set of unaltered digitized drop images were also selected for contact angle $\theta$ measurements (Test #18 to #21 in Table 1). The drop shapes used for this analysis are similar to the original profile of the drop with perfect reflection in Fig. 4. The computed $\theta$ from SIMO and OpenDrop are shown in red and blue, respectively (Fig. 7b). Snapshot of configuration panel of DropSnake used to predict $\theta$ is presented in Section S2 of supporting information. Please note that DropSnake does not assume drop shapes to be axisymmetric, exhibiting slight difference between the left and the right contact angle. Hence, the mean of the left and the right contact angle is presented in Fig. 7b for comparison with SIMO. For $\theta < 90°$, DropSnake consistently predicts $\theta$ slightly less than the actual value of $\theta$ while SIMO predicts it uniformly close to the actual $\theta$ with slight deviations. For $\theta > 90°$, DropSnake and SIMO predictions are consistent with actual $\theta$ till $\theta \sim 150°$. Upon further increase in $\theta$, computed $\theta$ slightly deviates from actual $\theta$ for both, the DropSnake and the SIMO.

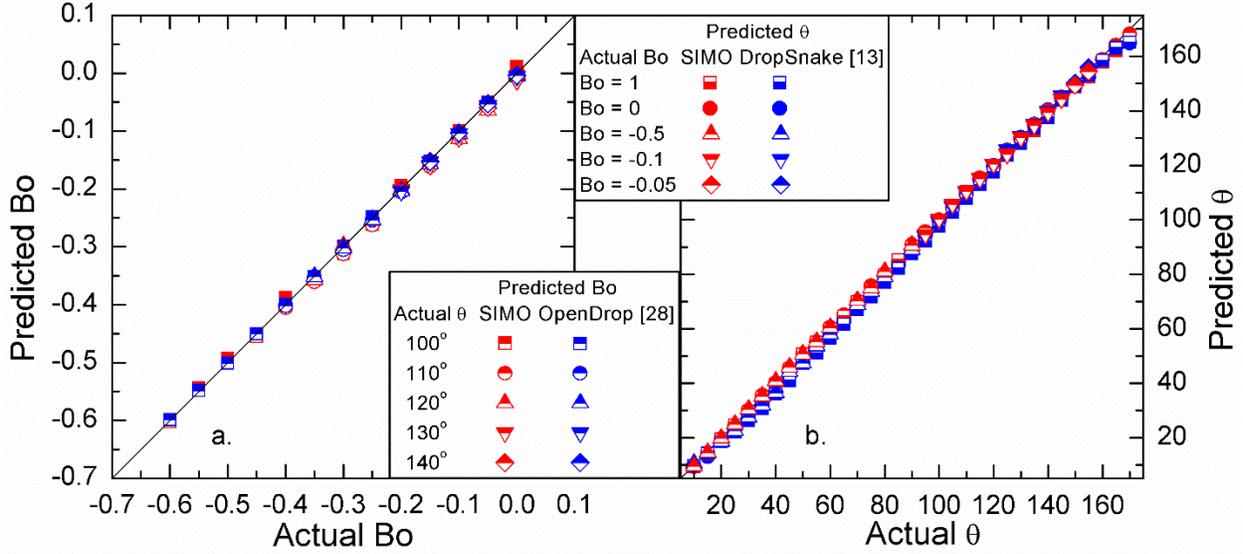

**Fig. 7.** (a) Predicted $Bo$ by SIMO (red) and OpenDrop (blue) [28] for digitized pendant drop images for Test #13 to #17. (b) Predicted $\theta$ by SIMO (red) and DropSnake (blue) [13] for Test #18 to #21.

We have established the accuracy in prediction of Bo and θ using SIMO for drop images generated from our droplet shape prediction algorithm [30]. In order to ascertain accuracy of SIMO for experimental drop images and to show that SIMO works equally well for experimental drop images, one complete cycle i.e., from low volume till detachment cycle of experimental drop images from our pendant drop tensiometry experiment discussed earlier are now analyzed using SIMO. It essentially employs the same procedure as adopted to analyze Fig. 2a with results in Fig. 3 for multiple images. Analyzed data of Fig. 2a is highlighted as filled blue star in Fig. 8. After computing Bond number by SIMO, surface tension was calculated employing Eqs. (1) and (2) and compared with results from similar experiments (Berry et.al. [28]) evaluated using OpenDrop in Fig. 8. Please note that here we determine Bond number as defined by Berry et.al. [28] ($Bo' = \rho g R_o^2 / \gamma$, where $R_o$ is radius of curvature of drop apex). Surface tension was plotted corresponding to the predicted values of $Bo'$ in Fig. 8. The scattered surface tension data for drops at low $Bo'$ is due to its fundamental physical limitation [24]. Drop shapes lying between the two vertical lines were chosen for surface tension calculation. The drop shape becomes unstable and the $Bo'$ reaches the critical value [36,37] at higher $Bo'$ beyond the vertical line. Mean value of surface tension evaluated using SIMO was obtained as 72.75 mN/m with a standard deviation of 0.31 for drop shapes lying between these two lines. The mean surface tension value is equal to the reported surface tension value of water ($72.75 \pm 0.36$ mN/m [34]) at 20 °C. Results from experiments by Berry et.al. [28] evaluated using OpenDrop show similar trend.

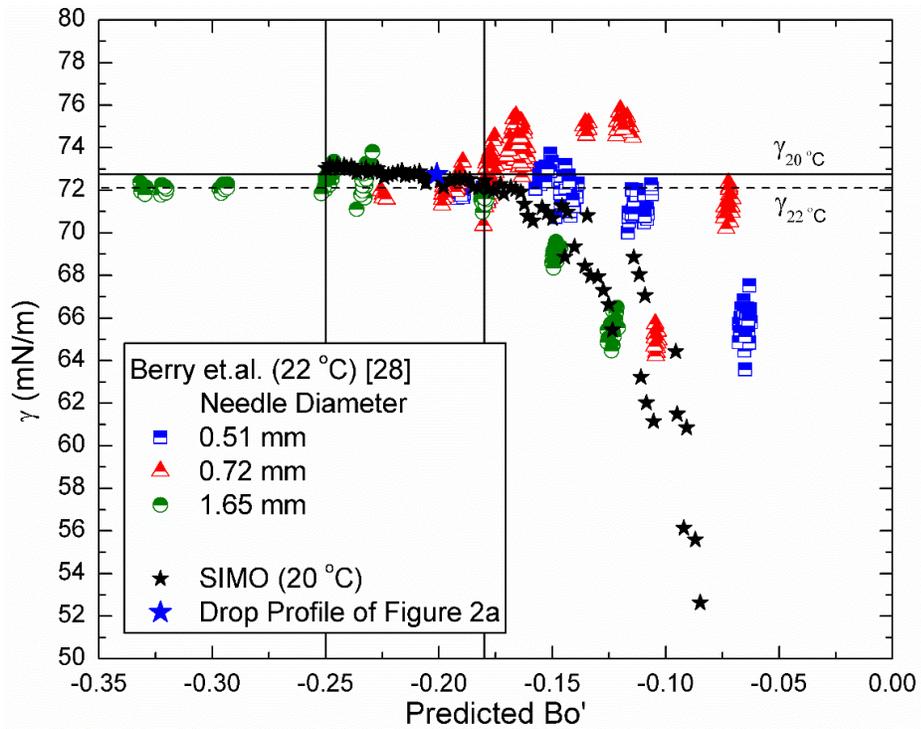

**Fig. 8.** Comparison of predicted surface tension from SIMO (black at 20 °C) and Berry et.al. [28] (OpenDrop – blue square, red triangle, green circle at 22 °C). The solid black horizontal line and dashed black horizontal line represent literature values of surface tension of water at 20 °C and 22 °C, respectively. Data represented by filled blue star symbol corresponds to the drop image analyzed in Fig. 2a.

## 4. Conclusion

We developed a Spline-based Interface Modeling and Optimization (SIMO) tool, a scheme to determine fluid-fluid interfacial tension and contact angle values from the axisymmetric profiles of sessile, pendant, and spherical drops. SIMO includes a geometric optimization algorithm which minimizes the thermodynamic free energy of drop by equalizing the equivalent curvature inside the drop to obtain equilibrium shapes. The expression for equivalent curvature is fundamentally an augmented Young-Laplace equation which encompasses the effects of surface tension and gravity. We show that the strategy is simple and robust and adequately predicts Bond numbers (and subsequently the surface tension) and contact angles across a range of values encountered in practical engineering applications. It can also evolve erroneous digitized profiles (errors due to image acquisition, edge detection, errors in the experimental setup, etc. in their interface) to their respective minimum energy equilibrium shapes (theoretical shape) for a precise estimation of surface tension and contact angle. The accuracy of the predicted values of surface tension and contacts angles from SIMO were shown to be comparable to existing open source alternatives such as OpenDrop [28] and DropSnake [13]. While the algorithm in its current form only includes Bond number as optimization parameter, we believe that further refinements can be achieved and the prediction accuracy can be improved by including volume and contact points as optimization parameters.

## Supporting Information

Section S1 includes coordinates of theoretical drop profile before and after pixelization. Snapshot of configuration panel of DropSnake is presented in Section S2. Section S3 describes the supporting videos.

## Notes

**Funding:** This research did not receive any specific grant from funding agencies in the public, commercial, or not-for-profit sectors.

**Competing Interest:** The authors declare no competing financial interest.

# Supporting Information: Spline-based Interface Modeling and Optimization (SIMO) for Surface Tension and Contact Angle Measurements

*Karan Jakhar[a,b], Ashesh Chattopadhyay[a,c], Atul Thakur[a], and Rishi Raj[a, 2]*

[a]*Department of Mechanical Engineering, Indian Institute of Technology Patna, Patna, Bihar 801103, India*

[b]*Currently with: J. Mike Walker'66 Department of Mechanical Engineering, Texas A&M University, College Station, TX 77843, USA*

[c]*Currently with: Department of Mechanical Engineering, Rice University, Houston, TX 77005,, USA*

## Supporting Information

### S1. Pixelized Drop Profile

**Table S1.** Coordinates of theoretical drop profile before and after pixelization.

| Theoretical Drop Profile | | Pixelized Drop Profile | |
|---|---|---|---|
| $x$ coordinate (mm) | $y$ coordinate (mm) | $x$ coordinate (mm) | $y$ coordinate (mm) |
| 0.4780 | 0.0000 | 0.4780 | 0.0000 |
| 0.5857 | 0.1138 | 0.5880 | 0.1052 |
| 0.6930 | 0.2279 | 0.6932 | 0.2199 |
| 0.7967 | 0.3453 | 0.7983 | 0.3442 |
| 0.8944 | 0.4678 | 0.8940 | 0.4589 |
| 0.9841 | 0.5963 | 0.9800 | 0.5928 |
| 1.0641 | 0.7309 | 1.0660 | 0.7266 |
| 1.1331 | 0.8715 | 1.1330 | 0.8700 |

[2] Corresponding Author. Telephone: +91 612 302 8166. E-mail address: rraj@iitp.ac.in {Rishi Raj}
Address: Room 113, Block III, Department of Mechanical Engineering, Indian Institute of Technology Patna, Bihta, Bihar 801103, India

| | | | |
|---|---|---|---|
| 1.1898 | 1.0176 | 1.1903 | 1.0135 |
| 1.2330 | 1.1682 | 1.2286 | 1.1664 |
| 1.2616 | 1.3222 | 1.2573 | 1.3194 |
| 1.2749 | 1.4783 | 1.2764 | 1.4724 |
| 1.2721 | 1.6349 | 1.2764 | 1.6254 |
| 1.2527 | 1.7903 | 1.2573 | 1.7879 |
| 1.2165 | 1.9427 | 1.2190 | 1.9409 |
| 1.1635 | 2.0902 | 1.1617 | 2.0843 |
| 1.0941 | 2.2306 | 1.0947 | 2.2277 |
| 1.0087 | 2.3619 | 1.0087 | 2.3616 |
| 0.9083 | 2.4822 | 0.9131 | 2.4763 |
| 0.7941 | 2.5894 | 0.7983 | 2.5815 |
| 0.6677 | 2.6819 | 0.6645 | 2.6771 |
| 0.5307 | 2.7579 | 0.5306 | 2.7536 |
| 0.3853 | 2.8162 | 0.3872 | 2.8109 |
| 0.2337 | 2.8557 | 0.2342 | 2.8492 |
| 0.0783 | 2.8756 | 0.0813 | 2.8683 |
| -0.0783 | 2.8756 | -0.0813 | 2.8683 |
| -0.2337 | 2.8557 | -0.2342 | 2.8492 |
| -0.3853 | 2.8162 | -0.3872 | 2.8109 |
| -0.5307 | 2.7579 | -0.5306 | 2.7536 |
| -0.6677 | 2.6819 | -0.6645 | 2.6771 |
| -0.7941 | 2.5894 | -0.7983 | 2.5815 |
| -0.9083 | 2.4822 | -0.9131 | 2.4763 |
| -1.0087 | 2.3619 | -1.0087 | 2.3616 |
| -1.0941 | 2.2306 | -1.0947 | 2.2277 |
| -1.1635 | 2.0902 | -1.1617 | 2.0843 |
| -1.2165 | 1.9427 | -1.2190 | 1.9409 |
| -1.2527 | 1.7903 | -1.2573 | 1.7879 |
| -1.2721 | 1.6349 | -1.2764 | 1.6254 |
| -1.2749 | 1.4783 | -1.2764 | 1.4724 |
| -1.2616 | 1.3222 | -1.2573 | 1.3194 |
| -1.2330 | 1.1682 | -1.2286 | 1.1664 |
| -1.1898 | 1.0176 | -1.1903 | 1.0135 |
| -1.1331 | 0.8715 | -1.1330 | 0.8700 |
| -1.0641 | 0.7309 | -1.0660 | 0.7266 |
| -0.9841 | 0.5963 | -0.9800 | 0.5928 |
| -0.8944 | 0.4678 | -0.8940 | 0.4589 |
| -0.7967 | 0.3453 | -0.7983 | 0.3442 |
| -0.6930 | 0.2279 | -0.6932 | 0.2199 |
| -0.5857 | 0.1138 | -0.5880 | 0.1052 |
| -0.4780 | 0.0000 | -0.4780 | 0.0000 |

**S2. Configuration panel of DropSnake**

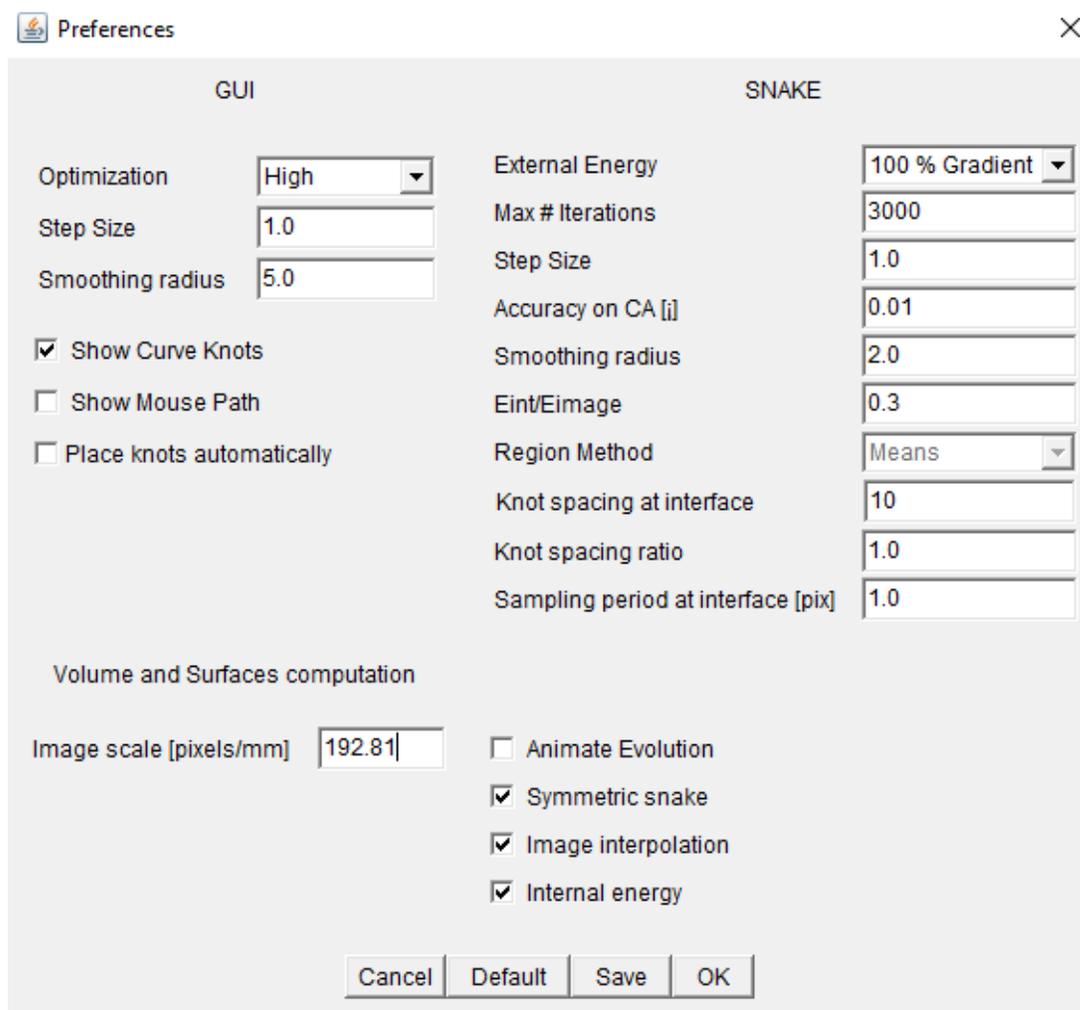

**Fig. S1.** Snapshot of configuration panel of DropSnake [1] used to predict $\theta$.

**S3. Videos**

**Video S1.** Evolution of initialized/inaccurate drop profile with non-uniform pressure distribution across liquid interface to the final equilibrium profile with uniform pressure distribution by SIMO corresponding to the assumed value of $Bo = -0.15$. (a) Profile of the evolving drop. (b) Corresponding values of $\chi_{mean}$ and $\chi_{eq}$ along the liquid-vapor interface Here $u = 0$ and $u = 1$ represent the start and the end points (contact points) of the parametrized spline, respectively. (c) $\sigma_{\chi_{eq}}$ for the evolving drop profile.

**Video S2.** Evolution of initialized/inaccurate drop profile with non-uniform pressure distribution across liquid interface to the final equilibrium profile with uniform pressure distribution by SIMO corresponding to the assumed value of $Bo = -0.237$. (a) Profile of the evolving drop. (b) Corresponding values of $\chi_{mean}$ and $\chi_{eq}$ along the liquid-vapor interface Here $u = 0$ and $u = 1$ represent the start and the end points (contact points) of the parametrized spline, respectively. (c) $\sigma_{\chi_{eq}}$ for the evolving drop profile.